\documentclass{cimento}

\newcommand{\mean}[1]{\left\langle{#1}\right\rangle}

\newcommand{\conds}{\langle \bar s s \rangle}
\newcommand{\condl}{\mean{\bar q q}_l}

\usepackage{cite}
\usepackage{graphicx,graphics,color}% Include figure files

\title{Thermal hadron resonances and Ward identities: results for the QCD phase diagram }
\author{A.~G\'omez Nicola\from{ins:x}\ETC,
J.~Ruiz de Elvira\from{ins:x},
        \atque
A.~Vioque-Rodr\'iguez\from{ins:x}
}
\instlist{\inst{ins:x} Departamento de F\'isica Te\'orica and
IPARCOS, Universidad Complutense de Madrid, Plaza de las Ciencias 1, 28040 Madrid, Spain
}
\begin{document}
\maketitle

\begin{abstract}
We review recent work regarding the role of light scalar resonances at finite temperature for chiral symmetry and $U(1)_A$ restoration. The results obtained are based on unitarized Chiral Perturbation Theory and Ward Identities and are directly connected with presently open problems within the QCD phase diagram. 
\end{abstract}

\section{Motivation}

Hadronic matter under extreme conditions such as finite temperature and/or chemical potentials in connection with the QCD phase diagram has been thoroughly studied lately, through theoretical models based on effective theories including Chiral Perturbation Theory (ChPT)~\cite{Pisarski:1983ms,Bochkarev:1995gi,Gerber:1988tt,Schenk:1993ru,Rapp:1999ej,Dobado:2002xf,Nicola:2013vma,GomezNicola:2016ssy,GomezNicola:2017bhm,Nicola:2018vug,Ferreres-Sole:2018djq,Nicola:2020smo}, lattice simulations \cite{Aoki:2009sc,Bazavov:2011nk,Bazavov:2014pvz,Cossu:2013uua,Buchoff:2013nra,Brandt:2016daq,Tomiya:2016jwr,Ratti:2018ksb,Bazavov:2019lgz,Bazavov:2019www,Aoki:2020noz} and experimental results from Heavy-Ion Collisions \cite{Adamczyk:2017iwn,Andronic:2017pug}. Although most of the relevant Physics within the QCD phase diagram has been understood, there are still some open problems \cite{Nicola:2020smo,Ratti:2018ksb,Bazavov:2019lgz}. Among them,  we will be interested here in the relation between the  Chiral Symmetry Restoration (CSR) and $U(1)_A$ asymptotic restoration, which affects the very nature of the transition and has many theoretical and phenomenological consequences depending on how effective $U(1)_A$ restoration is at the QCD chiral transition temperature $T_c\simeq$ 155 MeV~\cite{Pisarski:1983ms,GomezNicola:2017bhm,Nicola:2018vug,Nicola:2020smo,Cossu:2013uua,Buchoff:2013nra,Brandt:2016daq,Tomiya:2016jwr,Ishii:2016dln,Shuryak:1993ee,Kapusta:1995ww,Cohen:1996ng,Meggiolaro:2013swa,Pelissetto:2013hqa,Dick:2015twa,GomezNicola:2019myi}. In addition, as discussed below, an important role in this context is played by thermal resonances, i.e., resonances generated within the heavy-ion or thermal environment, whose spectral properties can be considerably modified due to the interactions among particles in the thermal bath.

The main observables regarding CSR are the light quark condensate $\condl=\langle \bar u u + \bar d d\rangle\equiv \langle \bar q_l q_l \rangle$, the order parameter of the chiral transition,  and the light scalar susceptibility $\chi_S$, defined as the two-point light $\bar q q$ correlator at vanishing four-momentum. On the one hand, the inflection point of $\condl$ at the crossover-like transition temperature $T_c$ signals CSR for physical masses, while the condensate vanishes at a lower $T_c$ in the light chiral limit $m_u=m_d=0$, where the transition is of second order, and $\chi_S$ peaks for physical masses and diverges in the chiral limit at $T_c$~\cite{Aoki:2009sc,Bazavov:2011nk,Ding:2019prx}. On the other hand, screening masses and susceptibilities in different channels provide also relevant information, as they would become degenerate at chiral and/or $U(1)_A$ restoration. Such chiral or $U(1)_A$ partners are particularly useful regarding the effective theory description, as commented below. Consider for instance the lightest meson states in the scalar/pseudoscalar nonets for isospin $I=0,1/2,1$, whose corresponding quark bilinears degenerate  as
%\footnote{Here, $\delta$ stands for the state corresponding to the $a_0(980)$ while $\sigma_l,\eta_l$ would correspond to the light part of the $f_0(500)/f_0(980)$ and $\eta/\eta'$ respectively.} 
\begin{eqnarray*}
\pi^a=i\bar q_l\gamma_5\tau^a q_l &
\stackrel{SU(2)_A}{
%\substack{\longleftrightarrow\\\mbox{\footnotesize $O(4)$ partners}}
\longleftrightarrow}
%\stackrel{SU(2)_A}{\longleftrightarrow}}
& \sigma_l=\bar q_l q_l,
\\
\qquad \qquad {\big\updownarrow \mbox{\tiny $U(1)_A$}} & &\qquad {\big\updownarrow \mbox{\tiny $U(1)_A$}}
%\end{eqnarray*}
%\hspace{-2cm}
%\begin{eqnarray*}
\\
\delta^a =\bar q_l \tau^a q_l  &
\stackrel{SU(2)_A}{
%\substack{\longleftrightarrow\\\mbox{\footnotesize $O(4)$ partners}}
\longleftrightarrow}
%\stackrel{SU(2)_A}{\longleftrightarrow}}
&  \eta_l=i\bar q_l \gamma_5 q_l ,
\\
\\
%\end{eqnarray*}
%\begin{eqnarray*}
 K^a=i\bar q  \gamma_5 \lambda^a q &
%\hspace*{0.2cm}
%{\color{red} \stackrel{SU(2)_A,U(1)_A}{\substack{\longleftrightarrow\\ \mbox{\scriptsize {\color{red} $U(1)_A$} }}}}
\stackrel{SU(2)_A,U(1)_A}{\longleftrightarrow}
%\hspace*{0.2cm}
&
\kappa^a=\bar q \lambda^a q,
%\color{black}
%\hspace*{0.1cm} \footnotesize
%$O(4)$ {\color{red} \&} $U(1)_A$ partners
\end{eqnarray*}
where $\delta$ stands for the state corresponding to the $a_0(980)$ and $\sigma_l, \eta_l$ would correspond to the light-quark part of the $f_0(500)/f_0(980)$ and $\eta/\eta'$, respectively. 
The states connected through a $SU(2)_A$ transformation are chiral partners, i.e., those whose associated screening masses and susceptibilities should become degenerate for $T\geq T_c$,  while those connected by an $U(1)_A$ transformation would degenerate asymptotically for large temperatures.

Lattice simulations yield the following degeneration pattern: for  $N_f=2+1$ light flavors. i.e., including strangeness, chiral partners degenerate reasonably around $T_c$ while $U(1)_A$ partner degeneration takes place at a significantly higher temperature, typically around  $1.3\,T_c$ from which the system would reach a $O(4)\times U(1)_A$ restoration pattern~\cite{Buchoff:2013nra,Bazavov:2019www}. The case of the $K-\kappa$ sector deserves some additional comments: while that pair degenerates both by  $SU(2)_A$ and $U(1)_A$ transformations, lattice results indicate that both symmetries need to be restored for $K-\kappa$ degeneration. In the particular $N_f=2$ case, results are qualitatively different, being compatible with a strong $U(1)_A$ restoration at $T_c$ in the chiral limit~\cite{Cossu:2013uua,Brandt:2016daq,Tomiya:2016jwr,Bazavov:2019www}. An important deal of theoretical work in the last few years has been devoted to explaining the differences in the behavior of these two cases, as summarized below.

\section{Ward Identities and partner degeneration}

\label{sec:WI} 

Ward Identities derived formally from the QCD generating functional have proven to be a very valuable tool to explore the degeneration of chiral and $U(1)_A$ partners \cite{Nicola:2013vma,GomezNicola:2016ssy,GomezNicola:2017bhm,Nicola:2018vug,Nicola:2020smo,GomezNicola:2019myi,GomezNicola:2020qxo,GomezNicola:2023rqi}. Let us summarize here some relevant results in those works. The following WI is particularly revealing in that respect:
\begin{equation}
\chi_P^{\eta_l\eta_s}(T)=-\frac{m_l}{2m_s} \left[\chi_P^{\pi\pi} (T)-\chi_P^{\eta_l\eta_l}(T)\right]=-\frac{2}{m_l m_s}\chi_{top}(T),
\label{WIls}
\end{equation}
where $\chi_P^{\phi_a\phi_b}(T)$ are pseudoscalar susceptibilities corresponding to $\langle \phi_a\phi_b\rangle$ correlators, $\eta_s=i\bar s \gamma_5 s$ and $\chi_{top}$ is the topological susceptibility corresponding to the $U(1)_A$ gluon anomaly operator. Now, one can perform a $SU(2)_A$ transformation rotating the bilinears $\eta_l\rightarrow \delta$ keeping $\eta_s$ invariant and since  $\langle\delta\eta_s\rangle$ is a scalar-pseudoscalar correlator, it vanishes by parity. Therefore,  $\chi_P^{\eta_l\eta_s}$ vanishes if chiral symmetry is completely restored and, since $\chi_P^{\pi\pi} -\chi_P^{\eta_l\eta_l}$ in (\ref{WIls}) vanishes when {\em both} chiral and $U(1)_A$ are restored, the above WI is consistent with $U(1)_A$ restoration at exact CSR. Such a result is compatible with the $N_f=2$ lattice results in the chiral limit since in that degenerate limit CSR is exact at at $T_c$. Nevertheless, the effect of both strangeness and a nonzero physical quark mass will generate a gap between chiral and $U(1)_A$ restoration, as observed in $N_f=2+1$ simulations. Such behavior has been checked in a $U(3)$ ChPT calculation, showing in particular that the transition temperatures of chiral and $U(1)_A$ degeneration of different partners tend to coincide in the light chiral limit \cite{Nicola:2018vug}. In addition, (\ref{WIls})  establishes  that $\chi_{top}(T)$ behaves  as an order parameter of   $O(4)\times U(1)_A$ restoration, as confirmed by theoretical analyses \cite{GomezNicola:2019myi,Azcoiti:2016zbi,diCortona:2015ldu}  and lattice  simulations \cite{Bonati:2015vqz}.   Two additional important WIs are those pertaining to the $I=1/2$ sector since they clarify the role of strangeness:
\begin{equation}
\chi_P^K(T)=-\frac{\condl (T)+2\conds (T)}{m_l + m_s}, \quad 
%\label{wiK}\\
\chi_S^\kappa (T)=\frac{\condl (T)-2\conds (T)}{m_s-m_l}.
  \label{wikappa} 
\end{equation}

\begin{figure}[h]
%% Use the relevant command for your figure-insertion program
%% to insert the figure file.
%% For example, with the option graphics use
\centering
%\hspace*{-0.35cm}
\resizebox{0.55\columnwidth}{!}{
\includegraphics{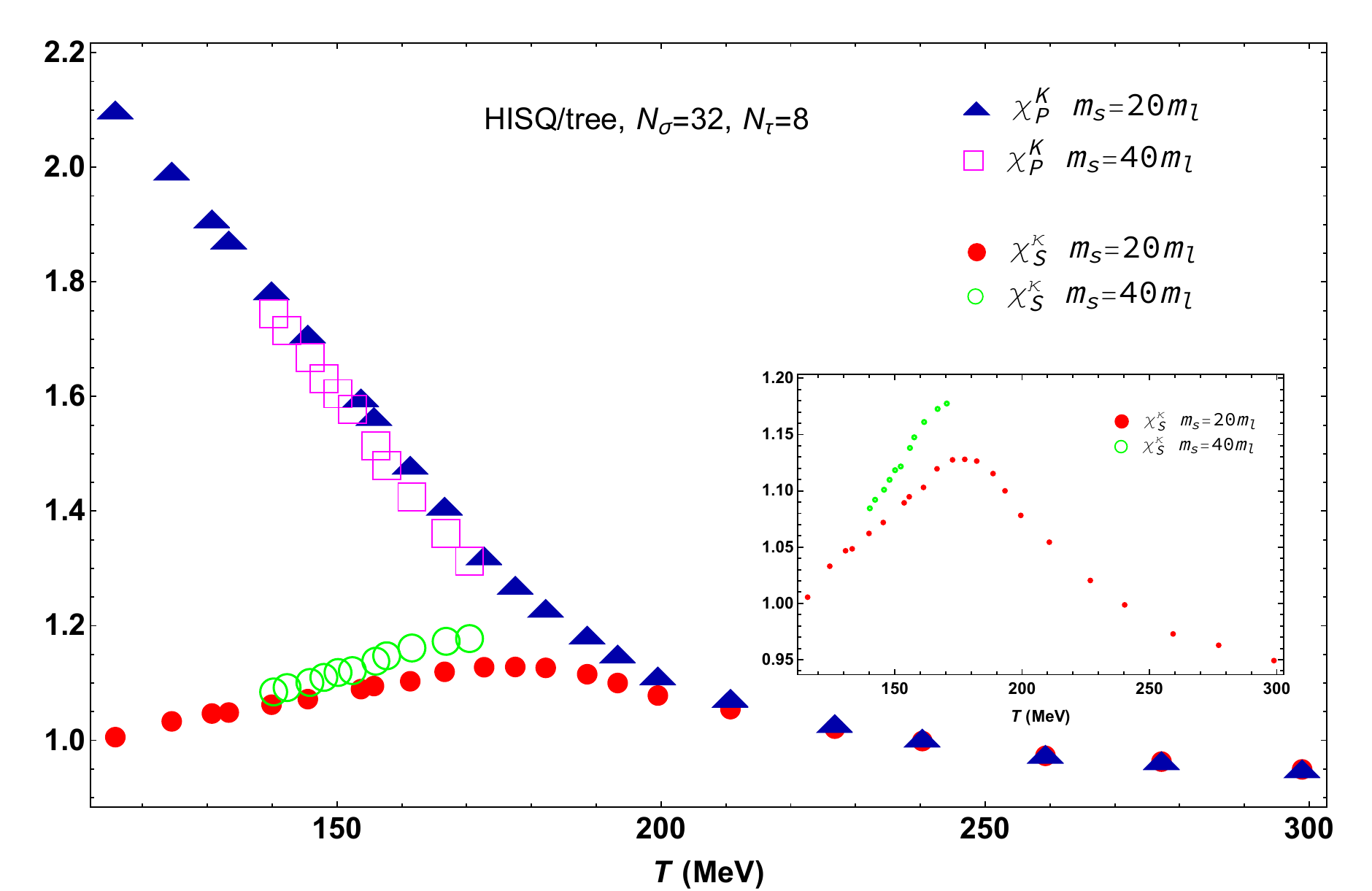}
}
%}
%\hspace*{-0.35cm}
%\resizebox{0.53\columnwidth}{!}{
%\includegraphics{susunit4.pdf}}
\caption{Susceptibilities in the $K-\kappa$ sector  extracted from the WIs in (\ref{wikappa}) and the quark condensate lattice values in \cite{Bazavov:2011nk,Bazavov:2014pvz}.}
\label{fig:Kkappa}       % Give a unique label
\end{figure}

Thus, taking into account that, due to CSR,   $\condl (T)$ decreases with $T$ much more abruptly than $\conds (T)$, the identities~(\ref{wikappa}) imply, on the one hand, that $\chi_P^K(T)$ should decrease monotonically,
while $\chi_S^\kappa (T)$ peaks above $T_c$ \cite{GomezNicola:2020qxo}. The behavior of $\chi_S^\kappa (T)$  below such peak is controlled by CSR, the slope increasing with stronger CSR effects, like reducing the pion mass. Above the peak, $U(1)_A$ restoration operates, driving  $\chi_S^\kappa$ towards degeneration with $\chi_P^K$. In Fig. \ref{fig:Kkappa}, we plot the lattice $\chi^{K,\kappa}$ susceptibilities ``reconstructed" from the WI (\ref{wikappa})  from condensate data, since there are no direct lattice determinations of susceptibilities available in those channels.  One can clearly observe the behavior we have just commented on. 

\section{Thermal Resonances}

The role of thermal effects in light hadron resonances has been recently proved to be crucial to explain many properties of the QCD phase diagram, in particular those related to CSR and the $U(1)_A$ symmetry that we are interested in here. Thus, in  \cite{Dobado:2002xf,Nicola:2013vma,Ferreres-Sole:2018djq} it has been shown that the thermal $f_0(500)$ generated from the Inverse Amplitude Method (IAM) unitarization of $\pi\pi$ scattering at finite temperature
\cite{GomezNicola:2002tn} dominates the scalar susceptibility through the saturation relation
\begin{equation}
\frac{\chi_S (T)}{\chi_S (0)}=\frac{M_S^2(0)}{M_S^2(T)},
\label{unitsus}
\end{equation}
where $M_S^2(T)$ plays the role of the $f_0(500)$ thermal mass, degenerating with the pion mass as $T$ approaches CSR \cite{Nicola:2020smo}, and is given by $M_S^2 =M_p^2-\Gamma_p^2/4$, being $s_p=(M_p-i\Gamma_p/2)^2$ the pole of the resonance in the $s$-complex plane in the second Riemann sheet across the thermal unitarity cut. It is worth pointing out that at finite temperature,  scattering of the incoming and outgoing particles with thermal bath components modifies the $T=0$ unitarity relation in two main aspects \cite{Nicola:2013vma,GomezNicola:2023rqi,GomezNicola:2002tn}, namely, phase space factors are corrected with Bose-Einstein thermal distribution functions and new Landau-like pure thermal cuts appear for the case of unequal masses. 
\begin{figure}[h]
%% Use the relevant command for your figure-insertion program
%% to insert the figure file.
%% For example, with the option graphics use
\centering
%\hspace*{-0.35cm}
\resizebox{0.5\columnwidth}{!}{
\includegraphics{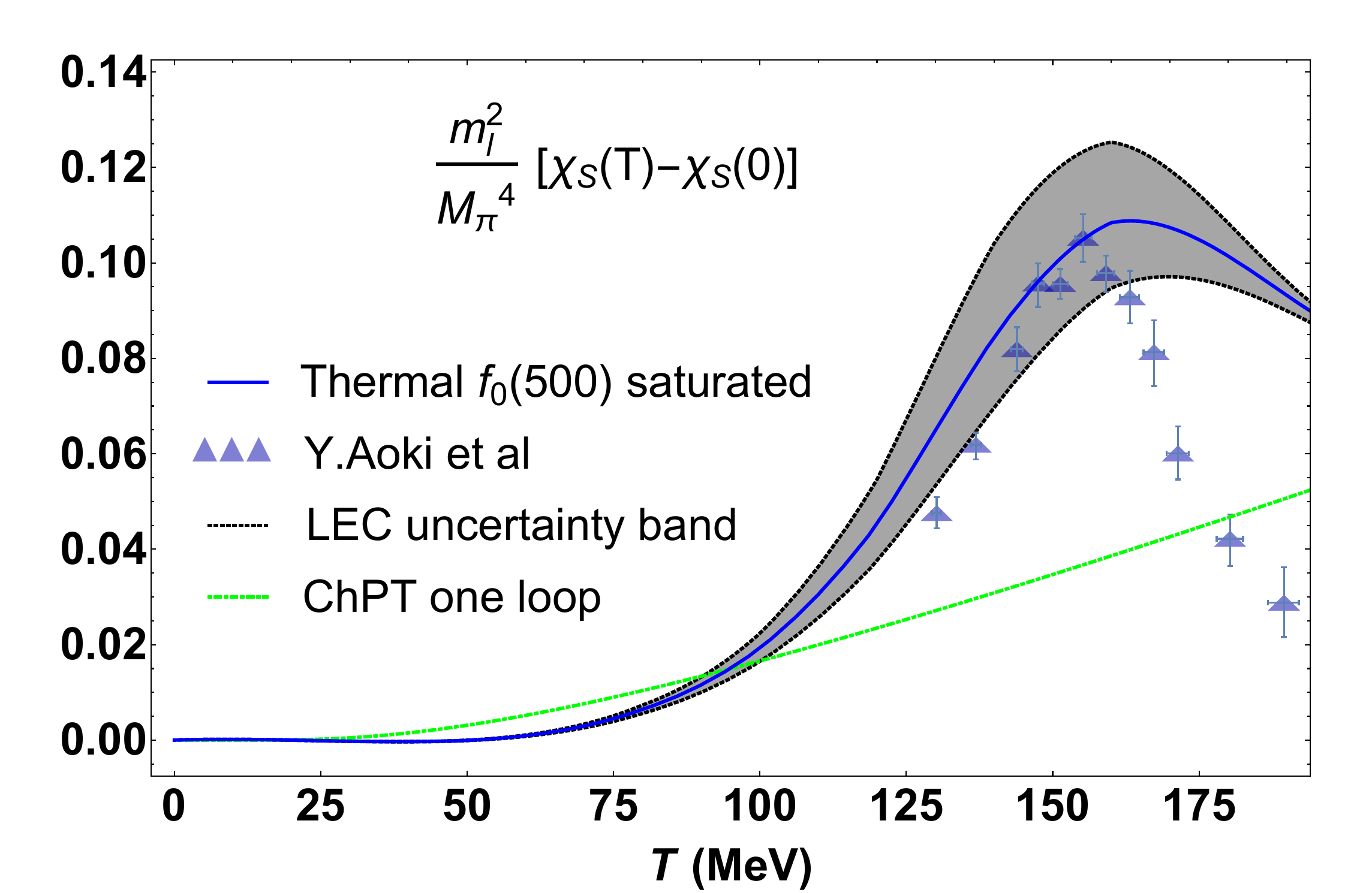}
}
\resizebox{0.48\columnwidth}{!}{
\includegraphics{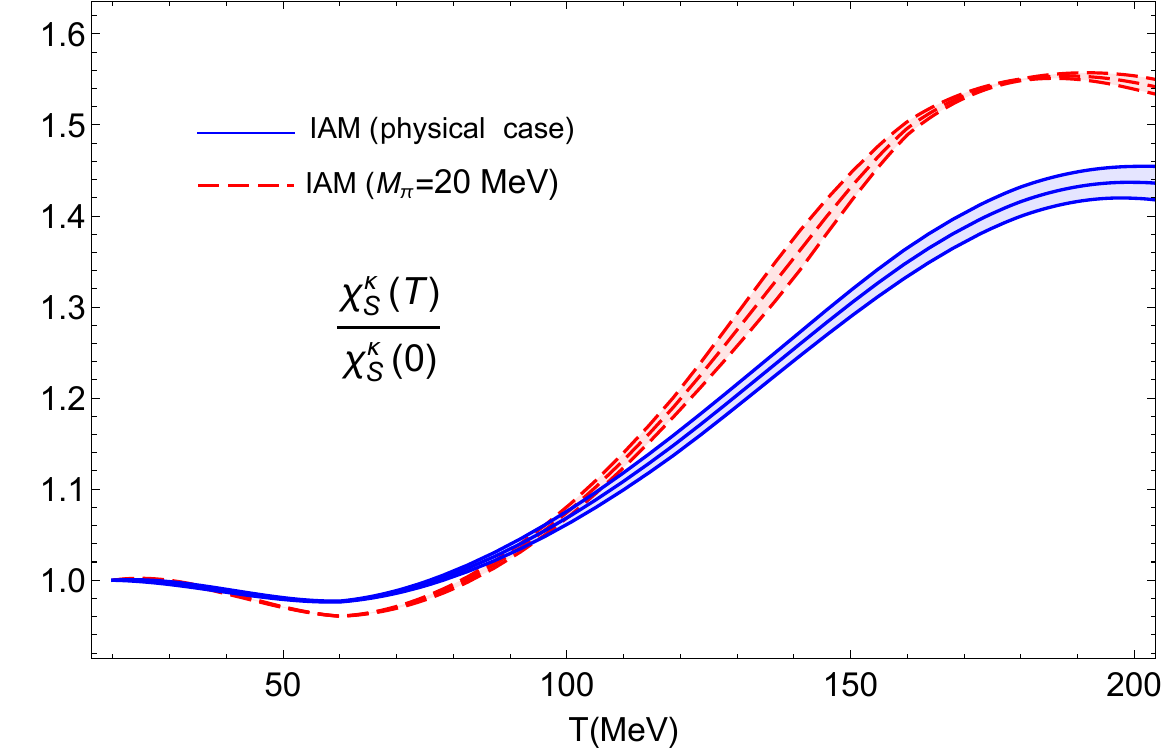}
}
%}
%\hspace*{-0.35cm}
%\resizebox{0.53\columnwidth}{!}{
%\includegraphics{susunit4.pdf}}
\caption{Unitarized scalar  susceptibilties $\chi_S$ (left) and $\chi_S^\kappa$ (right) within the IAM thermal resonance approach.  In the left panel, the lattice points come from \cite{Aoki:2009sc}, and  $\chi_S(0)$ is chosen to match the ChPT value in \cite{GomezNicola:2012uc}. The bands  correspond to the LECs uncertainties.} 
\label{fig:susunit}       % Give a unique label
\end{figure}

In Figure \ref{fig:susunit} we see that the unitarized saturated scalar susceptibility reproduces the expected transition peak and is compatible with lattice results below and around the peak within the uncertainty range of the Low Energy Constants (LEC) of the ChPT NLO Lagrangian, unlike the purely perturbative ChPT result \cite{GomezNicola:2012uc}, which is monotonically increasing and lies well above the lattice points for temperatures approaching the transition. The same approach has been followed for the thermal $K_0^* (700)$ or $\kappa$ resonance, generated from $\pi K$ scattering at finite temperature  \cite{GomezNicola:2020qxo,GomezNicola:2023rqi}. In this case, the unitarized saturation approach reproduces the expected peak for $\chi_S^\kappa$ 
 explained in section \ref{sec:WI}. In Fig. \ref{fig:susunit}, we show the result for $\chi_S^\kappa$ within this approach, confirming the expected behavior towards the light chiral limit, seen also in Fig.~\ref{fig:Kkappa}. In the $SU(3)$ limit $m_s\rightarrow m_l$, one finds $\chi_S^\kappa\rightarrow \chi_S$ as expected.  Thermal resonances in the vector channel $\rho (770)$, $K^* (890)$ have also been studied, with results compatible with heavy-ion phenomenology~\cite{Dobado:2002xf,GomezNicola:2023rqi}.
\section{Conclusions}
Thermal resonances and Ward Identities play a crucial role in understanding key properties of the QCD phase diagram, like those related to chiral and $U(1)_A$ restoration reviewed here. The lightest scalar thermal resonances saturate the corresponding susceptibilities in those channels, giving rise to results compatible with lattice and WI when those resonances are generated from unitarized meson-meson scattering. 
\acknowledgments
Work partially supported by research contracts  PID2019-106080GB-C21 and PID2022-136510NB-C31 (Spanish ``Ministerio de Ciencia e Innovaci\'on") and by the European Union Horizon 2020 research and innovation program under grant agreement No 824093. JRE acknowledges support from the Swiss National Science Foundation, proj.No.\ PZ00P2\_174228 and the Ram\'on y Cajal program (RYC2019-027605-I) of the Spanish MINECO. A. V-R acknowledges support from a fellowship of the UCM predoctoral program.

\end{document}